# N-Queens-based Algorithm for Moving Object Detection in Distributed Wireless Sensor Networks

Biljana Stojkoska, Danco Davcev, Trajkovik Vladimir
*Faculty of Electrical Engineering and Information Technologies, Karposh 2, b.b  Skopje,
Republic of Macedonia*
biles@fiet.ukim.edu.mk, etfdav@feit.ukim.edu.mk, trvlado@feit.ukim.edu.mk

**Abstract.** *The main constraint of wireless sensor networks (WSN) in enabling wireless image communication is the high energy requirement, which may exceed even the future capabilities of battery technologies. In this paper we have shown that this bottleneck can be overcome by developing local in-network image processing algorithm that offers optimal energy consumption. Our algorithm is very suitable for intruder detection applications. Each node is responsible for processing the image captured by the video sensor, which consists of NxN blocks. If an intruder is detected in the monitoring region, the node will transmit the image for further processing. Otherwise, the node takes no action. Results provided from our experiments show that our algorithm is better than the traditional moving object detection techniques by a factor of (N/2) in terms of energy savings.*

**Keywords.** Wireless Sensor Networks, Multimedia, Image Processing, Motion Detection.

## 1. Introduction

In recent years, there has been a huge advancement in wireless sensor computing technology [3]. This has led to the production of wireless sensors not only capable of observing and reporting physical phenomena, but also of accomplishing other operations, like data processing and communication. The sensors are organized in a network and they communicate by exchanging information using radio modules. They are actually responsible for the first stages of the processing hierarchy. After taking samples from the environment they sense (light level, air temperature, humidity etc.) they can process data or exchange it. All useful packets are sent to the sink node. This node is usually wirily connected with a server which stores the data. Server (base station) might be the final destination of the data or might act as a hub from where the data is sent to users over the wired network (Fig. 1).

The data is not directly sent to the sink as sensors nodes have short range of radio communication while being deployed in a vast region. There are a lot of multi-hops routing protocols that offer optimal communication cost. Each sensor sends data to its closest neighbor responsible for retransmitting the packets [9].

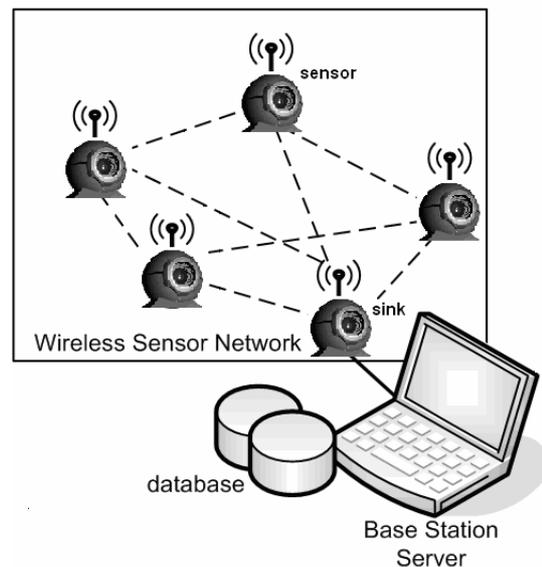

**Figure 1. Architecture of the system**

With the growth of mobile internet and multimedia services, wireless image sensors are becoming suitable for applications which require more detailed analysis of the environment, like security and video surveillance. The ability of WSNs to provide support for video streaming is restricted due to the bandwidth and power limitations of the nodes. Video streams unlike data streams are larger in size and require various guarantees for bandwidth, delay and jitter from the network [13]. A WSN is labored to provide enough end-to-end bandwidth for such communications. In addition, sensor nodes running on small batteries are power-starved, as battery technology has not been keeping pace with microelectronic technology. By using the up-to-date WSN routing protocols, this operation will result in unaccepted occupation of the bandwidth for a long period of time which will reduces the efficiency of the system.

In such a scenario, maintaining the lifetime of the nodes and enabling the video streams to travel over the network are two diametrically opposite requirements.

In this paper we describe fast in-network image processing algorithm for surveillance application, which is especially suitable for intruder detection applications.

The rest of the paper is organized as follows. In the next section, the relevant work related to the present mote platforms for camera sensor network is discussed. Section three of this paper describes the traditional techniques for motion detection and explains the framework we propose for WSN image processing. The forth section explains our N-Queens based algorithm for local in-network image processing. The fifth section gives the results provided from our experiments and demonstrates the effectiveness of the proposed algorithm in minimizing energy consumption for wireless sensor communication. Finally, we conclude this paper in section six.

## 2. Related work

Unless WSN are touted as a popular solution in surveillance systems, little work has been done in integrating image sensors into wireless node.

Mica2 platform provides an easy way to add the capability of capturing images just by plugging a module onto the mote. The camera and the mote are interfaced using an intermediate board providing an interface to combine them [5]. MICA motes can be purchased from Crossbow technology, Inc. They are equipped with a radio module, 8-bit 4-7 MHz processor, memory, two AA batteries and a suite of sensors [6].

Cyclops [10] is a smart vision sensor board that can be attached as an external sensor to a mote, such as the one from the Mica family (Fig.2 left). Cyclops consists of a micro-controller (MCU), a complex programmable logic device (CPLD), an external SRAM, an external Flash and an imager that offers a limited resolution of the images (128x128).

Weeble node architecture is another video platform [11] that represents an integration of low cost, wide angle camera with Mica2 motes (Fig.2 right). This node is capable of taking images at a maximum resolution of 640x480 and additionally can self-correct for arbitrary orientation.

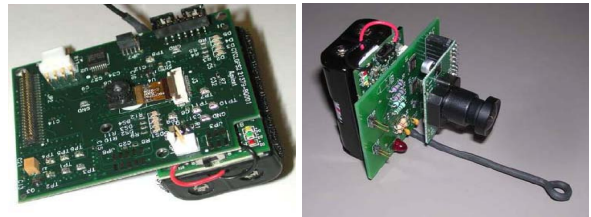

**Figure 2. Cyclops [10] (left) Weeble Architecture [11] (right)**

SensEye[12] is a camera sensor network for video surveillance based on three tier architecture. Tier 1 consists of Cyclops nodes that wake up periodically, obtain an image of the environment and process the image to detect presence of new objects using a simple frame differencing mechanism. An image of the background is stored in the memory and is used for framing differences for each captured image. The frame difference is passed through a simple threshold-based noise filter to get a cleaned foreground image. The number of foreground pixels together with a thresholding mechanism is used to detect new objects in the environment. Higher tiers consist of more advanced motes with high-resolution cameras.

## 3. Framework for WSN Image Processing

Motion detection is the first building block in a typical surveillance application. The next stages (object tracking and higher level processing) heavily depend on the accuracy and robustness of the first step of moving object detection [1].

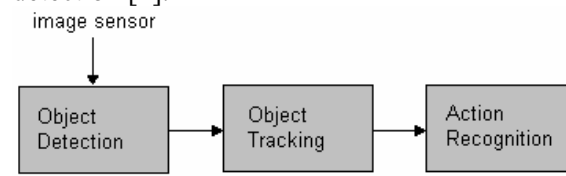

**Figure 3. Generic Framework for Image Processing algorithms**

In the traditional systems for video surveillance (Fig.3), the video sequence is captured by the image sensor and compressed using JPEG compression. The module for motion detection receives a wide-angle camera image as an input and computes the difference between consecutive images. When this difference is above the threshold, it announces the presence of a moving object. Motion detection is still considered as a difficult problem, due to dynamic changes in the nature, like sudden illuminations, weather changes, waving trees, water etc.

Most of the algorithms that are proposed within the last years and solve the problem of extracting moving targets from a video stream, basically, can be divided into three main groups: temporal differencing, background subtraction and optical flow [1]. The optical flow approach is a very complex and time consuming technique that is inappropriate for such real-time applications and is not subject of our interest [1].

Temporal differencing is highly adaptive to dynamic environments, since it uses the pixel-wise difference between two or three consecutive frames in video imagery. This method does not perform well in extracting all relevant feature pixels, especially when the foreground object stops moving, moves slowly or its texture is uniform. This algorithm is also known as temporal thresholding.

The most commonly used algorithms for motion detection are based on background subtraction. This method is very sensitive to dynamic environment but provides complete feature data. The detection is achieved by subtracting the background from the current frame in all regions where they match. The background image is created in an initialize phase by averaging images over time. In order to adapt to the current environmental conditions, the reference background is updated with new images over time.

When modeling the background, there are usually two common problems we are faced with:
- the presence of foreground objects during the process of modeling the background;
- sudden variations in illumination conditions;

Modeling of the background is usually carried out at pixel level, where pixels could be collected in a number of blocks. The intensity or color of the pixels is the commonly used feature.

Apart from traditional systems for video surveillance, in WSN applications data can not be proceeded in traditional fashion. For a simple example, consider an application for intruder detection. A network of wireless sensor nodes has been deployed to monitor an environment. Each node in the network is equipped with a video sensor that periodically captures images from the scene, compresses them using JPEG compression and transmits them to the sink which is directly connected to the central server where the further processing is done.

Such high-rate, real-time imaging applications pose significant design challenge, as they require strict end-to-end delay, bandwidth, and jitter guarantees:
- In the most multi-hops routing protocols, each node sends the packets to its parent. The nodes that are highest in the routing tree (and closest to the sink), will have the most data to transmit and will quickly exhaust their available energy. In this situation, the useful information could not reach the sink any more.
- Since multi-hop communication tends to generate more interference, delay, and both higher packet loss and error during transmission, this operation will result in unaccepted occupation of the bandwidth for a long period of time. Interference and high packet loss rate affect the bandwidth and delay values of the route. This can not satisfy the strict quality of service requirements, such as sustaining transmission of high quality data at a high bit-rate.

To overcome the bottlenecks of wireless image communication we propose two-level architecture for video surveillance systems. The nodes represent the first level since they are responsible for image capturing and initial frame processing. Image sensor nodes must process the images in-node to extract relevant features before transmission. Nodes locally apply the algorithms for motion detection, and if the provided results are decided to be useful, the picture is compressed and sent to the sink. Various compression standards have been analyzed for WSN efficient power management, but JPEG is considered as the most appropriate technique [4][8].

The second level is represented by the server (base station) which is responsible for further analysis (object tracking and action recognition) as shown on Fig. 4.

This architecture suits well for large scale networks used for applications such as military tracking or intruder detection. The nodes are scattered randomly in a hostile and unattended environment. After deployment, they would form a topology map and would start to monitor the area for any movement or unusual activity. If a particular node detects a moving object, it could compress the current frame (or only the critical block of the frame) and send it over the network to the base station.

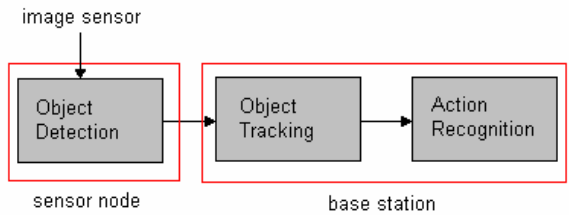

**Figure 4. Image Processing Framework in large scale Wireless Sensor Networks**

## 4. In-network motion detection

WSN framework for image processing discussed in previous section reflects only on reducing network traffic, therefore minimizing communicational cost and increasing lifetime of the network. Another issue that arises in wireless image sensor networks is the complex image processing algorithms, which are both memory and computational intensive. Data encoding/decoding involves significant processing, much more than the processing required for other sensed data. Further, aggregation at the intermediate nodes requires complicated algorithms and higher processing power.

These requirements make image processing a challenging problem, as sensor nodes are resource constraint devices. Memory limitations, local computational energy cost and processor speed have to be considered in order to guarantee real-time response. To rise above these shortcomings, we propose a new fast algorithm for motion detection.

We were inspired by the famous problem of N-Queens distribution on a chess field. The queens are put on an N×N chessboard such that no two queens attack each other. This N-queens based paradigm was previously proposed for block matching techniques, as in [12] where N-Queen pixel decimation is used for fast motion estimation. We apply a similar idea as a technique for motion detection.

### 4.1. N-queens based algorithm for motion detection

The trajectories of the moving objects can have any direction, column, row, or diagonal in the image. If the input frame is divided into N columns and N rows, and N queens are placed such that:

- none of said N blocks occupies a position in the same row as any other one of said N blocks;
- none of said N blocks occupies a position in same column as any other one of said N blocks; and,
- none of said N blocks occupies a position in the same diagonal as any other one of said N blocks;

it is very possible that most of the moving object trajectories can cross at least one queen. It is because we expect that trajectories in the nature are usually continual. It is not very common to have zig-zag trajectories.

By applying modified background subtraction technique, we propose to reduce the image analyze. Instead of comparing all $N^2$ blocks from background and current image, we can compare only N blocks that belong to the queens, hoping that moving object will cross at least one queen.

We made a small experiment. At the beginning 64 queens were placed on 32 x 32 grid cells. Generating random trajectories, for each we check if it passes though at least one queen. We had positive results in 75% of 100 samples. As can be seen from the picture (Fig.5 (a)), the trajectories with red color are not captured. To decrease this possibility, queens can be applied on more positions. One way is by reflecting the initial N queens across central vertical line (Fig.5 (b)). This distribution we call double N queens distribution. Other combinations are also possible. For double N queens distribution the experiment was positive in 95%.

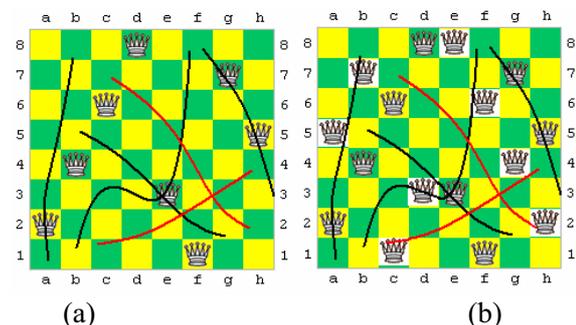

(a)　　　　　　　　(b)

**Figure 5. N-Queens (a) and Double N-Queens (b) distribution**

Motivated by this analysis, we propose the following algorithm:

Step 1. capture first image;
Step 2. store first image in background frame buffer;

Step 3. forever:
- capture new image
- store new image in the input frame buffer
- divide the input frame into N columns and N rows;
- choose N referent blocks that correspond to the positions of N-Queens problem distribution;
- compare each of the N blocks from the input frame with the same block from the background frame;
- if the difference of at least one block is above the threshold, send the frame to the sink and go to step 3;
- if the difference is below the threshold, update the background frame and go to step 3;

The main challenge here is modeling the background. For simplicity we pronounce the first frame as a background, although it can be created by averaging images over time. To respond to the environmental changes, if the block difference is below the threshold, we move the background frame slightly in the direction of the current frame. In other words, we update pixels values toward new values. It has to be mentioned that choosing the threshold is not only application-specific, but also depends on the specification of the environment.

The pseudo code of the algorithm is shown below:

```
read_background_image;
while (true){
   i:=1;
   read_new_image;
   while ( i<=2*N) {
    q_position =next(q_position);
    s = n_f[q_position]-bg_f [q_position];

    if (abs(s)>threshold){
        send_alarm;
        break;
     }
    else {
        if (s>0)
          bg_f [q_position]++;
        else
          bg_f [q_position]--;
     }
    i++;
   }
}
```

The function `read_background_image` reads the background image and stores it in one dimensional array `bg_f`. The function `read_new_image` reads the new image and stores it in one dimensional array `n_f`. The function `send_alarm` sends the new image to the sink. The function `next` returns the position of the next queen which has to be processed.

## 5. Evaluation

First we tested our N-Queens based algorithm with double N-Queens distribution on video streams taken from web cameras (320 × 240 pixels camera image). We implemented image processing routines within a high level processing language. To compare the performances, we processed the frames using traditional techniques based on background subtraction on a frame-by-frame basis (left pictures on Fig. 6) and using our N-Queens based algorithm (right pictures on Fig. 6). Using 8 × 8 matrix representation, pixels are collected in a number of blocks (each block consists of 64 pixels). The initial image was converted into 40 x 30 pixels image, where each pixel represents the average of one block. We compared 1200 blocks for background subtraction and 64 blocks for N-Queens based algorithm. These 64 blocks satisfy the double N-Queens distribution.

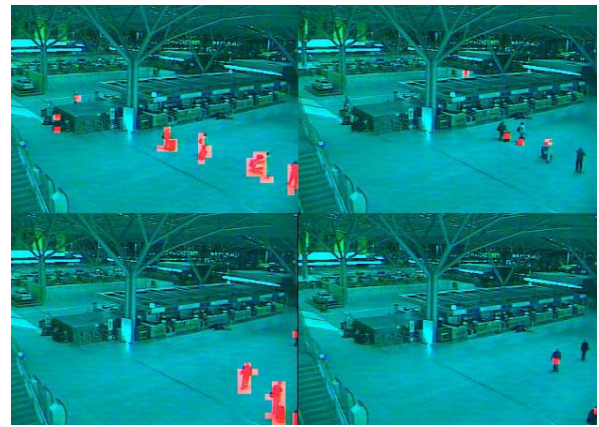

**Figure 6. Motion detection using background subtraction algorithm (left) and N Queens based algorithm (right)**

To analyze the performance achievement for Mica2 platform, simulation software was implemented in NesC, a structured, component-based language, very similar to the programming language C, and the testbed was simulated using Avrora [5] simulator.

We measured the energy consumption for processing one image sample. The power needed to capture the image is not included in our experiments. The results are shown in the table below:

|  | Energy Usage (mJ) | |
| --- | --- | --- |
| Algorithm | 128x128 | 176x255 |
| Background Subtraction | 23.19 | 63.78 |
| N-Queens-based | 2.9 | 5.79 |

As can be seen from the results, our N-Queens based algorithm is better than the traditional background subtraction technique by a factor of (N/2) in terms of energy savings. As N-Queens based algorithm for moving object detection is a heuristic method, we suggest an adaptive approach which is combination of our algorithm and the traditional background subtraction. For example, if we capture images at every 5 seconds, then four images can be processed with our algorithm, and the fifth image can be processed with traditional background subtraction method. The number of images processed with our algorithm is application specific. If we are interested in detecting intruders, we can compute which is the common speed of the people passing cross. If the time needed for the people to cross the whole region is 20 seconds, we can use our algorithm for 19 image samples, and the traditional method for one image sample. Such an adaptive approach should increase the robustness of the particular application.

## 6. Conclusion

In this paper, we presented a new WSN image processing framework. In our approach, each node in the network is responsible for the first stage of the processing. If an intruder is detected, the image is transmitted to the sink. In order to minimize the energy consumption, we propose an in-network image processing algorithm based on the famous N-queens distribution of the blocks. The results we have presented show that our algorithm consumes (N/2) -times less energy for local image processing than traditional techniques.

If combined with traditional techniques, this approach can perform significant improvement of the network performances.